\documentclass[aps,
pra,
twocolumn,
superscriptaddress,
nofootinbib,
longbibliography]{revtex4-2}

\usepackage{amsmath,amssymb,amsthm,mathtools}
\usepackage{graphicx}
\usepackage{physics}
\usepackage{mathrsfs}
\usepackage{bm}
\usepackage{float}
\usepackage[utf8]{inputenc}

\DeclareMathAlphabet\mathbfcal{OMS}{cmsy}{b}{n}

\newcommand{\be}{\begin{equation}}
\newcommand{\ee}{\end{equation}}

\begin{document}

\title{Non-Associativity Induced Modifications of Open-System Quantum Dynamics: General Master Equation and a Two-Qubit Ising Case Study}

\author{Ekin S{\i}la Y\"or\"uk}
\email{eyoruk13@ku.edu.tr}
\affiliation{Department of Physics, College of Sciences, Ko\c c University, 34450 Sar\i yer, Istanbul, Türkiye}

\author{\"Ozg\"ur E. M\"ustecapl\i o\u glu}
\email{omustecaplioglu@ku.edu.tr}
\affiliation{Department of Physics, College of Sciences, Ko\c c University, 34450 Sar\i yer, Istanbul, Türkiye}
\affiliation{TÜBITAK Research Institute for Fundamental Sciences (TBEA), 41470 Gebze, Türkiye}

\author{Zafer Gedik}
\email{gedik@sabanciuniv.edu}
\affiliation{Faculty of Engineering and Natural Sciences,
Sabanci University, 34956 Tuzla, Istanbul, Türkiye}

\date{\today}

\begin{abstract}
Nonassociative deformations of phase-space structures arise naturally
in the presence of magnetic charge, where the Jacobi identity for
momentum components fails and the corresponding Moyal product becomes
nonassociative.
While such structures are well understood at the level of
single-particle kinematics, their implications for open-system quantum
dynamics remain largely unexplored.
Here we derive a Born--Markov master equation for a system coupled
to a bath when the underlying operator product is weakly
nonassociative. The deformation enters through associators appearing
in the second-order kernel, while pairwise operator products and
dissipators retain their standard form. The resulting correction is
dispersive and modifies the Liouville--von Neumann part of the
generator without introducing additional dissipative channels.
We then embed this structure into a two-qubit transverse-field Ising
model using a Stratonovich--Weyl representation and an Ising-aligned
twisted Poisson structure. In the zero-temperature limit, the nonassociative terms produce a
nonlinear correction in which the instantaneous population imbalance
of each qubit feeds back into the dynamics as a state-dependent
longitudinal field.
Numerical simulations in the weak-coupling regime, where the
Born--Markov derivation is quantitatively controlled, show that
increasing the nonassociativity parameter suppresses steady-state
entanglement by up to $59\%$, reduces purity, and increases entropy,
while leaving the relaxation timescale set by the dissipative rates
unchanged.
These results demonstrate that weak nonassociativity manifests as a
coherent, population-dependent deformation of open-system dynamics
rather than an additional dissipative mechanism.
\end{abstract}

\maketitle


\section{Introduction}
\label{sec:intro}

Noncommutativity is central to quantum theory and underlies many
operational advantages exploited in quantum technologies, including
precision metrology, quantum information processing, and engineered
light--matter interactions.
Because the commutator encodes the departure from classical physics at
the algebraic level, it is natural to ask whether other algebraic
deformations may influence quantum dynamics.
In particular, weakening associativity introduces an additional
structural modification that is independent of noncommutativity and
may lead to qualitatively new dynamical effects.

Nonassociativity arises in several physical contexts, including
momentum-space structures in magnetic monopole backgrounds \cite{Preskill1984}, twisted
Poisson geometries with nonvanishing Jacobiators, and field-theoretic
constructions with three-form fluxes
\cite{Szabo2019IntroNonAssocPhysics}, \cite{Szabo2018MagneticMonopoles},
\cite{schupp2024algebraic}, \cite{Jackiw1985ThreeCocycle},
\cite{Guneydin1985MagneticCharge}, \cite{Heninger2020HamiltonianMonopole}.
In such settings the operator product fails to satisfy the Jacobi
identity, and associators provide an independent measure of algebraic
deformation.
These ideas have been explored primarily at the level of kinematics,
including uncertainty relations and ground-state structure in monopole
backgrounds \cite{Bojowald2015States}, \cite{Bojowald2015Testing},
\cite{Bojowald2018SmallMagnetic}, \cite{Bojowald2021GroundState}.
Their implications for open quantum systems, however, remain largely
unexplored.

In open-system dynamics, nested commutators naturally appear in
the Born--Markov expansion. If the underlying operator algebra is
weakly nonassociative, these triple products acquire additional
contributions governed by associators. Importantly, this deformation
affects only the Liouvillian structure of the generator, while
pairwise operator products and dissipators retain their conventional
form. The resulting correction is therefore coherent and dispersive
rather than dissipative.

To address this problem, we derive a general Born--Markov master
equation for a system coupled linearly to a stationary reservoir when
the operator product is nonassociative
\cite{ottinger2010nonlinear}, \cite{dann2018time}.
Deviations from associativity appear at second order in the system--bath
coupling through operator associators weighted by bath correlation
functions.
The resulting master equation preserves trace and Hermiticity and
reduces to the standard Gorini--Kossakowski--Lindblad--Sudarshan (GKLS)
generator \cite{Gorini1976}, \cite{Lindblad1976} in the associative
limit.

The nonassociative contribution modifies the Liouville--von Neumann
term and generates a state-dependent coherent correction, while the
dissipative structure remains unchanged. This separation allows one to
interpret weak nonassociativity as a population-dependent deformation
of the coherent dynamics within an otherwise standard open-system
framework.

To illustrate the qualitative consequences of the theory, we analyze a
two-qubit transverse-field Ising model subject to local
amplitude-damping channels and a monopole-inspired nonassociative
deformation aligned with the Ising axis.
This minimal setting allows us to examine how the deformation modifies
relaxation pathways, coherence dynamics, steady-state properties, and
bipartite correlations.

Numerical simulations show that nonassociativity primarily reshapes
steady states and suppresses entanglement, while leaving the
characteristic Lindblad relaxation timescale largely unchanged. These
features indicate that weak nonassociativity acts as a coherent
deformation of open-system dynamics rather than an additional
dissipative mechanism.

The remainder of this article is organized as follows.
In Sec.~II we summarize the twisted Poisson and Jacobiator structures
that motivate the nonassociative deformation. Sec.~III introduces the
Stratonovich--Weyl correspondence and the qubit embedding. Sec.~IV
presents the general nonassociative Born--Markov master equation.
Sec.~V introduces the two-qubit TFIM, and Sec.~VI reports numerical
results. Sec.~VII concludes.


\section{Monopole-induced nonassociativity in phase space}
\label{sec:monopole}

We begin by recalling the standard phase-space realization of
nonassociativity in the presence of a magnetic charge density \cite{Szabo2018MagneticMonopoles}, \cite{Jackiw1985ThreeCocycle}, \cite{Bojowald2015Testing}.
Consider a particle of electric charge \(q\) moving in a magnetic
field \(\mathbf B(\mathbf x)\).
In canonical coordinates \((x_i,p_j)\), the standard Poisson brackets
are deformed as
\begin{equation}
\{x_i,x_j\} = 0,
\ \
\{x_i,p_j\} = \delta_{ij},
\ \
\{p_i,p_j\} = q\, \epsilon_{ijk} B^k(\mathbf x),
\label{eq:magnetic-poisson}
\end{equation}
with \(i,j,k\in\{1,2,3\}\) and $\epsilon_{ijk}$ the Levi-Civita
tensor.

It is convenient to collect these brackets into an antisymmetric
bivector \(\theta_{IJ}(z) = \{z_I,z_J\}\) with
\(z^I \in \{x_1,x_2,x_3,p_1,p_2,p_3\}\).
The nonvanishing components are
\begin{equation}
\theta_{x_i p_j} = \delta_{ij},
\qquad
\theta_{p_i p_j} = q\, \epsilon_{ijk} B^k(\mathbf x).
\end{equation}
If the magnetic field is divergence-free, \(\nabla \cdot \mathbf B = 0\),
the above defines a genuine Poisson structure.
However, in the presence of a magnetic charge density
\(\rho_m = \nabla \cdot \mathbf B \neq 0\),
the Jacobi identity for the momentum components fails.
Explicitly,
\begin{align}
J(p_i,p_j,p_k)
& := \{p_i,\{p_j,p_k\}\}  \nonumber \\ &+ \{p_j,\{p_k,p_i\}\} + \{p_k,\{p_i,p_j\}\} \nonumber \\
& = q\, \epsilon_{ijk}\, \nabla \cdot \mathbf B.
\label{eq:jacobiator}
\end{align}

This violation of the Jacobi identity can be equivalently encoded by
an antisymmetric rank-3 tensor
\begin{equation}
H_{ijk}(\mathbf x) = -\, q\, \epsilon_{ijk} \, \nabla \cdot \mathbf B,
\end{equation}
which can be regarded as the components of a closed 3-form
\begin{equation}
H = \tfrac{1}{6} H_{ijk}\, dx_i \wedge dx_j \wedge dx_k.
\end{equation}
A bivector with a nonvanishing Jacobiator of this form is called a
twisted Poisson structure \cite{szabo2019quantization}.

Given a Poisson bivector \(\theta_{IJ}(z)\), deformation quantization
yields the Kontsevich star product \cite{kontsevich2003deformation}
\begin{align}
(f \star g)(z)
= & f g
+ \frac{i\hbar}{2}\,\theta_{IJ}\,\partial_I f\,\partial_J g \nonumber \\
& - \frac{\hbar^2}{8}\,\theta_{IJ}\theta_{KL}\,
\partial_I \partial_K f\,\partial_J \partial_L g
+ \mathcal O(\hbar^3).
\end{align}
Failure of associativity is captured by the associator
\begin{equation}
[f,g,h]_\star := (f\star g)\star h - f\star(g\star h).
\end{equation}
At leading order in~$\hbar$ one finds
\begin{equation}
[f,g,h]_\star \;=\; \frac{(i\hbar)^3}{6}\,J(f,g,h) \;+\; \mathcal O(\hbar^4),
\end{equation}
where the Jacobiator is written directly in Poisson-bracket form,
\begin{equation}
J(f,g,h) \;:=\; \{f,\{g,h\}\}+\{g,\{h,f\}\}+\{h,\{f,g\}\}.
\end{equation}
For the monopole-twisted structure introduced above, only the momentum
block contributes, while all other components vanish.
Substituting \eqref{eq:jacobiator} into the leading term gives, for
arbitrary smooth $f,g,h$,
\begin{equation}
[f,g,h]_\star
\;=\; \frac{(i\hbar)^3}{6}\,q\,\epsilon_{ijk}\,(\nabla\!\cdot\!\mathbf B)\,
(\partial_{p_i} f)\,(\partial_{p_j} g)\,(\partial_{p_k} h)
\;+\; \mathcal O(\hbar^4).
\end{equation}
Consequently, when the magnetic field is divergence-free
($\nabla\!\cdot\!\mathbf B=0$), the local star product is associative,
whereas a nonzero magnetic charge density produces a controlled,
leading nonassociative correction.

Using the identity involving the Levi--Civita symbol, this can be
expressed in terms of a triple scalar product as
\begin{equation}
[f,g,h]_\star
= \frac{(i\hbar)^3}{6}\,q\,(\nabla\!\cdot\!\mathbf B)\,
\big( \nabla_p f \times \nabla_p g \big)\cdot \nabla_p h
+ \mathcal O(\hbar^4).
\end{equation}
Thus, in flat three-dimensional phase space, the leading nonassociative
contribution to the star product is proportional to the divergence of
the magnetic field and to the oriented volume spanned by the gradients
of $f,g,h$ in momentum space.

This construction provides a concrete realization of weak
nonassociativity at the level of phase-space functions. In what
follows, we use this structure only as a guiding example: the
resulting associator will later be embedded into qubit degrees of
freedom through the Stratonovich--Weyl correspondence, leading to a
nonassociative deformation of open-system dynamics.


\section{Stratonovich--Weyl correspondence and qubit embedding}
\label{sec:SW}

To connect the phase-space description of nonassociativity with
finite-dimensional quantum systems, we employ the Stratonovich--Weyl
(SW) correspondence \cite{Stratonovich1957},
\cite{VarillyGraciaBondia1989}, which admits a systematic
group-theoretical formulation in terms of $SU(2)$ coherent states on
the Bloch sphere. This mapping allows operator products to be
represented as star products between functions on $S^2$, providing a
natural setting to introduce controlled nonassociative deformations.
We refer the reader to \cite{Klimov2009} for a comprehensive treatment
of this correspondence in the context of atom-field interactions, and
summarize here only the elements essential to the present construction.

This mapping assigns to each operator $A$ acting on a qubit Hilbert
space a real function $\tilde{A}(\mathbf{n})$ defined on the Bloch
sphere $S^2$ according to
\begin{equation}
\tilde{A}(\mathbf{n}) := \mathrm{Tr}\!\left[A\,\Delta(\mathbf{n})\right],
\end{equation}
where $\Delta(\mathbf{n})$ is the SW kernel.
Concretely, for a single qubit the SW kernel
\begin{equation}
\Delta_{1/2}(\mathbf n)
=\tfrac12\!\left(\mathbb{I}+\sqrt3\,\mathbf n\!\cdot\!\boldsymbol{\sigma}\right),
\qquad \mathbf n\in S^2,
\end{equation}
yields symbols $\tilde{A}(\mathbf n)=\mathrm{Tr}[A\,\Delta_{1/2}(\mathbf n)]$;
in particular $\widetilde{\sigma_i}(\mathbf n)=\sqrt3\,n_i$.
The map is invertible, preserves Hermiticity, and is covariant under
$SU(2)$ rotations. Most importantly, it satisfies traciality, ensuring
that operator averages are faithfully reproduced as integrals of their
symbols over the sphere.
Throughout this work, we use a tilde to denote SW symbols, so that,
for example, $\tilde{H}$ is the SW symbol of the Hamiltonian $H$.

For two qubits, the correspondence extends naturally to the product
space $S^2\times S^2$, with
\begin{equation}
\tilde{\mathcal{A}}(\mathbf{n}^{(1)},\mathbf{n}^{(2)}) = \mathrm{Tr}\!\left[\mathcal{A}\,\Delta_{1/2}(\mathbf{n}^{(1)})\otimes\Delta_{1/2}(\mathbf{n}^{(2)})\right].
\end{equation}
In that case we take the tensor kernel
$\Delta(\mathbf n^{(1)},\mathbf n^{(2)})
=\Delta_{1/2}(\mathbf n^{(1)})\otimes\Delta_{1/2}(\mathbf n^{(2)})$
and use $\boldsymbol{\sigma}^{(a)}:=\sqrt3\,\mathbf n^{(a)}$ as ambient
$\mathbb{R}^3$ coordinates for each site $a=1,2$.

In our representation, the operator product is replaced by a star
product between symbols.
Throughout this construction, nonassociativity is introduced at the
level of the effective star product for the symbols, and therefore
appears as a deformation of the dynamical generator rather than a
modification of pairwise operator products.
In the associative case, this star product reproduces the operator
algebra exactly.
In the presence of a magnetic charge density or a twisted Poisson
structure, however, the star product fails to be associative, with the
associator controlled by the underlying Jacobiator.

Since the effective degrees of freedom are the Bloch vectors, we
introduce a weak nonassociative deformation directly in the symbol
space by postulating a twisted Poisson structure in the $\sigma$
variables. This construction serves as an effective analogue of the
monopole-induced deformation discussed in Sec.~\ref{sec:monopole},
adapted to finite-dimensional spin degrees of freedom.
Concretely, we take a block-diagonal bivector on
$\mathbb{R}^3_{\sigma^{(1)}}\times \mathbb{R}^3_{\sigma^{(2)}}$,
\begin{equation}
\{f,g\}_\Pi \;=\;
\sum_{a=1}^2 F^{(a)}(\sigma^{(a)})\cdot
\big( \nabla_{\!\sigma^{(a)}} f \times \nabla_{\!\sigma^{(a)}} g \big),
\label{eq:twisted-bivector}
\end{equation}
where $F^{(a)}(\sigma^{(a)})$ denote the local coefficients of the
twisted Poisson bivector acting on the Bloch coordinates.
At leading order the Kontsevich associator reads

\begin{widetext}
\begin{equation}
[f,g,h]_\star \;=\; \sum_{a=1}^2
\left(-\,\frac{i\hbar^3}{6}\right)\,
\chi^{(a)}(\sigma^{(a)})
\big(\nabla_{\!\sigma^{(a)}} f \times \nabla_{\!\sigma^{(a)}} g\big)\!\cdot \nabla_{\!\sigma^{(a)}} h, \qquad
\chi^{(a)} := F^{(a)}\!\cdot(\nabla_{\!\sigma^{(a)}}\times F^{(a)}).
\label{eq:assoc-sigma}
\end{equation}
\end{widetext}

The explicit prefactor $-i\hbar^3/6$ guarantees that for real-valued
SW symbols the associator is anti-Hermitian.

It should be emphasized that the Stratonovich--Weyl correspondence by
itself does not generate nonassociativity.
In the magnetic-charge picture the Jacobiator arises from the twisted
Poisson structure of momentum space.
By contrast, in the two-qubit embedding there are no canonical momenta,
and hence no direct derivation of a Jacobiator.
To model an effective analogue of the monopole-induced deformation, we
postulate a twisted Poisson bivector
$\Pi=\Pi^{(1)}\oplus\Pi^{(2)}$ on the Bloch phase space as in
Eq.~\eqref{eq:twisted-bivector}.
This ansatz should be understood as an effective parametrization of
weak nonassociativity rather than a fundamental derivation. It
preserves Hermiticity and trace of the reduced dynamics, and is
compatible with the symmetry breaking of the Ising Hamiltonian, which
singles out the $z$ axis.

We therefore adopt $\chi^{(a)}(\sigma)=\kappa_a\sigma^{(a)}_z$ as
the simplest symmetry-compatible choice, defining an Ising-aligned
nonassociative deformation in the two-qubit setting. The real
parameters $\kappa_a$ quantify the strength of the deformation and
control the magnitude of the resulting nonlinear feedback in the
master equation derived below.


\section{Nonassociative Born--Markov master equation: general structure}
\label{sec:generalME}

We now summarize the general form of the nonassociative Born--Markov
master equation that will be applied to the two-qubit TFIM.
A detailed derivation is given in Appendix~\ref{app:NAderivation}.
The derivation follows the standard second-order perturbative expansion
in the system--bath coupling \cite{dann2018time}, \cite{BreuerPetruccione2002}, \cite{d2021self}, \cite{Trushechkin2021}, \cite{Breuer2016} generalized here to accommodate a weakly
nonassociative operator product. The resulting correction appears in
the Liouville--von Neumann part of the generator, while the dissipative
structure retains the conventional GKLS form.

We consider a physical scenario where a primary quantum system $S$,
whose evolution is of interest, is coupled to an auxiliary quantum
system $B$, referred to as the bath.
The total Hamiltonian is
\begin{equation}
\hat{H}(t) = \hat{H}_S + \hat{H}_B + \alpha \hat{H}_{SB},
\end{equation}
where $\alpha \in [0,1]$ is an auxiliary parameter, used for
perturbation series expansion in terms of the system-bath interaction.
The state of the composite system is given by the global density
operator $\hat{\rho}_{SB}(t)$, which obeys the Liouville--von Neumann
equation
\begin{equation}
\frac{d}{dt} \hat{\rho}_{SB}(t) = -\frac{i}{\hbar} \left[ \hat{H}_S + \hat{H}_B + \alpha \hat{H}_{SB}, \, \hat{\rho}_{SB}(t) \right].
\end{equation}
Here $[A,B]:=AB-BA$, where pairwise operator products retain their standard associative form; nonassociativity enters only through triple products appearing in the second-order kernel, captured by the associator terms $\mathcal{A}_j$.

Let $\mathcal{L}_0$ denote the adjoint Liouvillian with respect to
this product,
\begin{equation}
\mathcal{L}_0(X) := \frac{i}{\hbar}\,[H_S+H_B,\,X].
\end{equation}
We define the interaction picture by the (left) action of the
superoperator exponential
\begin{align}
A_I(t) := e^{\,t\mathcal{L}_0}\,&A,
\quad
\rho_I(t) := e^{\,t\mathcal{L}_0}\,\rho_{SB}(t), \nonumber \\
& H_I(t) := e^{\,t\mathcal{L}_0}\,H_{SB}.
\end{align}
This avoids any bracketing ambiguities that may arise from writing
$U(t) A U^{-1}(t)$ in a nonassociative algebra.
With this convention,
\begin{equation}
\frac{d}{dt} \hat{\rho}_I(t) = -\frac{i \alpha}{\hbar} \left[ \hat{H}_I(t), \, \hat{\rho}_I(t) \right].
\label{eq:von-Neu1}
\end{equation}
Integrating once and re-inserting yields the usual second-order
expansion, but with the underlying product nonassociative.

Up to the point where the Born approximation is introduced, the formal
steps mirror the associative case, and one arrives at the
familiar-looking expression
\begin{align}
&\frac{d}{dt} \hat{\rho}_S(t) \nonumber \\ = &-\frac{1}{\hbar^2} \int_0^\infty dt_1 \, \mathrm{Tr}_B \left\{ \left[ \hat{H}_I(t), \, \left[ \hat{H}_I(t_1), \, \hat{\rho}_S(t) \otimes \hat{\rho}_B \right] \right] \right\}.
\label{eq:von-neu-jacob}
\end{align}
However, the second-order kernel involves triple products (nested
commutators); when the underlying product is nonassociative, additional
terms appear that are captured by associators.
These additional terms arise from triple operator products and
therefore modify only the coherent part of the generator. The
dissipative structure originating from bath correlation functions
remains unchanged.

To quantify this, we introduce the Jacobiator and associator for a
nonassociative product:
\begin{equation}
\text{J}(A, B, C) = [A, [B, C]] + [C, [A, B]] + [B, [C, A]],
\label{jacobiator}
\end{equation}
and
\begin{equation}
[A,B,C]:=(AB)C-A(BC).
\end{equation}
Nonassociativity leads to the failure of the Jacobi identity,
\begin{align}
\text{J}(A, B, C)& = [A, C, B] + [B, A, C] + [C, B, A] \nonumber \\ &- [A, B, C] - [B, C, A] - [C, A, B]\neq 0.
\label{jabobiator-associator}
\end{align}
Using these relations inside the double commutator that appears in the
Born--Markov kernel, one finds that the nonassociative corrections can
be organized into six distinct associator structures built from the
system operators.
The resulting general master equation has the form

\begin{widetext}
\begin{align}
\frac{d\rho_S}{dt}
&= -\frac{i}{\hbar}[H_S,\rho_S]
+ \Gamma_{+}\!\left(S\rho_S S^\dagger-\tfrac{1}{2}\{S^\dagger S,\rho_S\}\right)
+ \Gamma_{-}\!\left(S^\dagger\rho_S S-\tfrac{1}{2}\{SS^\dagger,\rho_S\}\right)
\nonumber\\
&\quad +
\Big[
\Lambda_1\,[S^\dagger,S,\rho_S]
+\Lambda_2\,[S,S^\dagger,\rho_S]
+\Lambda_3\,[S,\rho_S,S^\dagger]
+\Lambda_4\,[S^\dagger,\rho_S,S]
+\Lambda_5\,[\rho_S,S,S^\dagger]
+\Lambda_6\,[\rho_S,S^\dagger,S]
\Big],
\label{eq:na-master}
\end{align}
\end{widetext}
where $S$ is the system coupling operator, $\Gamma_\pm$ are the
standard dissipative rates, and the $\Lambda_j$ are complex
coefficients determined by bath correlation functions and the
nonassociative deformation.
The six coefficients satisfy
\begin{subequations}\label{eq:LambdaChains}
\begin{align}
\Lambda_{1}=\Lambda_{3}=\Lambda_{6} &= K_{+}-K_{+}^{*} = 2i\,\mathrm{Im}\,K_{+}, \\
\Lambda_{2}=\Lambda_{4}=\Lambda_{5} &= K_{-}-K_{-}^{*} = 2i\,\mathrm{Im}\,K_{-},
\end{align}
\end{subequations}
with
\begin{equation}
K_\pm \;:=\; \int_0^\infty d\tau\, C_\pm(\tau)
\;=\; \frac{\Gamma_\pm + i\,\varepsilon_\pm}{2},
\end{equation}
and $C_\pm(\tau)$ the usual two-time bath correlators (see
Appendix~\ref{app:NAderivation}).

It is worth noting that the nonassociative coefficients
$\Lambda_j$ in Eq.~\eqref{eq:LambdaChains} depend only on
$\mathrm{Im}\,K_{\pm} = \varepsilon_{\pm}/2$, not on the dissipative
rates $\Gamma_{\pm}$.
Consequently, $\mathcal{N}[\rho_S]$ is entirely dispersive in
character: it does not open additional dissipative channels, and its
strength is controlled by the dispersive part of the bath response,
$\varepsilon_\pm$, in analogy with the Lamb-shift contribution of
standard open-system theory.
Furthermore, $K_+$ is complex for both fermionic and bosonic baths,
since the bath correlator $C_+(\tau)$ is a complex-valued function
whose time integral generically yields nonvanishing real and imaginary
parts \cite{Brasil2013}.
The real part $\mathrm{Re}\,K_+ = \Gamma_+/2$ governs the standard
dissipative (amplitude-damping) rate, while
$\mathrm{Im}\,K_+ = \varepsilon_+/2$ is the dispersive contribution.
Both are determined by the same spectral density $J(\omega)$ via
Eq.~\eqref{eq:GammaEps}, so the relative strength of the
nonassociative correction is set by the ratio
$\varepsilon_+/\Gamma_+$, which can in principle be tuned by
engineering $J(\omega)$.

The nonassociative correction $\,\mathcal N[\rho_S]\,$ provided by
the last line of Eq.~\eqref{eq:na-master} is trace-preserving and
Hermiticity-preserving.
Trace preservation follows from the 3-cyclic property of the trace
functional in the nonassociative setting, which makes all associator
contributions vanish under the trace.
Hermiticity preservation follows from the adjoint identity
$[A,B,C]^\dagger=-[C^\dagger,B^\dagger,A^\dagger]$ and the relations
\eqref{eq:LambdaChains}.
In the zero-temperature, vacuum-like case for a fermionic bath, one
has $K_-=0$, and the correction reduces to
\begin{equation}
\mathcal{N}[\rho_S]
= \big(K_+ - K_+^{\ast}\big)\,
\Big(
[S^\dagger,S,\rho_S] + [S,\rho_S,S^\dagger] + [\rho_S,S^\dagger,S]
\Big).
\label{eq:final-N-T0}
\end{equation}

The nonassociative correction $\mathcal{N}[\rho_S]$ in
Eq.~\eqref{eq:final-N-T0} is entirely dispersive: the prefactor
$K_+ - K_+^* = 2i\,\mathrm{Im}\,K_+$ is purely imaginary, so the
correction preserves Hermiticity and does not introduce additional
dissipation.
We note, however, that the three-associator combination does
\emph{not} reduce to a standard commutator $-i[H,\rho_S]$ for a fixed
Hermitian $H$; the reduction to a specific operator form depends on
the underlying nonassociative structure and is carried out explicitly
in Sec.~\ref{sec:model} via the Stratonovich--Weyl correspondence.
This interpretation is closely related to the effective-potential
approach in nonassociative quantum
mechanics~\cite{Bojowald2015Testing}, where nonassociativity generates
state-dependent corrections to the semiclassical dynamics.
Such effective nonlinear master equation results should be interpreted
as identifying observable signatures of nonassociativity rather than
as a definitive test of fundamental nonassociative quantum mechanics.
Because the reduced nonlinear generator can in principle be emulated
by associative feedback-based platforms, the observed dynamics would
constitute evidence for a nonassociative model only insofar as
competing associative explanations can be excluded.

The vanishing of $K_-$ at zero temperature holds for both fermionic
and bosonic baths.
For a fermionic bath $C_-(\tau) \propto n_F(\omega) \to 0$ as $T \to
0$, while for a bosonic bath in the vacuum state $\hat{\rho}_B =
|0\rangle\langle 0|$ one has $\hat{B}(t)|0\rangle = 0$, which makes
the absorption correlator $C_-(\tau) = \langle \hat{B}^\dagger(t)
\hat{B}(t')\rangle_B$ vanish identically \cite{Brasil2013}.
At finite temperature, however, $K_-$ becomes nonzero for both bath
types, and all six coefficients $\Lambda_j$ in
Eq.~\eqref{eq:na-master} contribute to the nonassociative correction.
The generator in Eq.~\eqref{eq:na-master} reduces to the standard GKLS
form \cite{Gorini1976}, \cite{Lindblad1976}, \cite{Vacchini2017} when all associators
vanish, recovering the associative Born--Markov limit
\cite{dann2018time}, \cite{BreuerPetruccione2002}, \cite{d2021self}, \cite{Trushechkin2021}, \cite{Breuer2016}.
In what follows, this general generator will be specialized to a
two-qubit TFIM and a specific nonassociative deformation encoded via
the SW correspondence.
Additional discussion of the complete positivity of the
general \emph{linear} generator Eq.~\eqref{eq:na-master} is given in
Appendix~\ref{app:CP}.
We note that this proof, which relies on the secular approximation
applied to the linear associator superoperators, does not extend
directly to the nonlinear reduced equation obtained after the
Stratonovich--Weyl mapping (Secs.~\ref{sec:model}--\ref{sec:results}).
Complete positivity for that nonlinear equation is instead verified
numerically, as described in Sec.~\ref{sec:results}.


\section{Two-qubit transverse-field Ising model and nonassociative feedback}
\label{sec:model}

We now specify the system Hamiltonian and the nonassociative
deformation in the two-qubit setting.
We consider a setup of two interacting qubits governed by the
transverse-field Ising model (TFIM), which provides the simplest
framework that incorporates frustration, quantum fluctuations, and
monopole-like excitations.
The Hamiltonian is chosen as
\begin{equation}
H_{\mathrm{TFIM}} = -J\, \sigma_z^{(1)} \otimes \sigma_z^{(2)}
- \frac{h_1}{2}\, \sigma_x^{(1)} \otimes \mathbb{I}
- \frac{h_2}{2}\, \mathbb{I} \otimes \sigma_x^{(2)} ,
\end{equation}
where $\sigma_\alpha^{(a)}$ ($\alpha=x,z$) are Pauli matrices acting
on qubit $a=1,2$.
The longitudinal Ising coupling $J$ enforces alignment along the $z$
axis and encodes a local ``two-in--two-out'' ice-rule constraint, while
the transverse fields $h_a$ introduce coherent spin flips that connect
different degenerate configurations of the ice manifold.
Excitations that violate the ice rule correspond to emergent magnetic
monopoles~\cite{Castelnovo2008}, and their dynamics are enabled by the
transverse terms.
In this way, the TFIM captures both the classical spin-ice structure
and the additional quantum dynamics that arise in qubit spin-ice
architectures.

We restrict to the two-qubit case as the simplest nontrivial
realization that retains the essential physics while allowing for
explicit analytic treatment of the nonassociative corrections; two-qubit open system dynamics in similar collision-model settings has been studied in \cite{Cakmak2017}.
The initial state is assumed to be uncorrelated,
\begin{equation}
\rho(0) = \rho^{(1)} \otimes \rho^{(2)}, \qquad
\rho^{(a)} = \tfrac{1}{2}\left(\mathbb{I} + \vec{r}^{(a)} \cdot \vec{\sigma}^{(a)}\right),
\end{equation}
where $\vec{r}^{(a)}$ denotes the Bloch vector of qubit $a$.

Dissipation and decoherence are introduced by coupling each qubit
locally to an external bath through the operators
\begin{equation}
S^{(a)} = g_a \, \sigma_-^{(a)}, \qquad
\sigma_- = \tfrac{1}{2}(\sigma_x - i\sigma_y),
\end{equation}
with real coupling constants $g_a$.
This choice realizes amplitude-damping channels and provides a natural
description of relaxation in superconducting qubits.
Our general formalism, however, is not restricted to a specific type
of bath \cite{BreuerPetruccione2002}: the master equation derived above
depends only on bath correlation functions and thus applies equally to
fermionic and bosonic environments.
For concreteness, we focus on the fermionic case in the derivation and
then adopt the zero-temperature limit.

To embed the monopole-like nonassociativity discussed in
Sec.~\ref{sec:SW} into the TFIM dynamics, we adopt an Ising-aligned
twist of the form $\chi^{(a)}(\sigma) = \kappa_a \sigma_z^{(a)}$ in
Eq.~\eqref{eq:assoc-sigma}.
This choice preserves Hermiticity and trace of the reduced dynamics
while embedding monopole-like signatures into the qubit degrees of
freedom.
The constants $\kappa_a$ quantify the strength of nonassociative
effects on each site and encode the analogue of an underlying magnetic
charge distribution.

For simplicity, the two-qubit state is taken uncorrelated,
\begin{equation}
\rho=\rho^{(1)}\otimes\rho^{(2)},\qquad
\rho^{(a)}=\tfrac12\big(I+\vec r^{\,(a)}\!\cdot\vec\sigma^{(a)}\big),
\end{equation}
so that its SW symbol is
\begin{equation}
\tilde\rho=\tfrac14\Big(I\otimes I+\sum_i r^{(1)}_i \sigma^{(1)}_i\!\otimes I+\sum_j r^{(2)}_j I\!\otimes\sigma^{(2)}_j\Big),
\end{equation}
and
\begin{equation}
\nabla_{\sigma^{(1)}}\tilde\rho=\tfrac14\vec r^{\,(1)},\qquad
\nabla_{\sigma^{(2)}}\tilde\rho=\tfrac14\vec r^{\,(2)}.
\end{equation}

For a fixed site $a$ one has
\begin{align}
& \nabla_{\sigma^{(a)}} S^{(a)}=g_a\Big(\tfrac12,\,-\tfrac{i}{2},\,0\Big), \nonumber \\
& \nabla_{\sigma^{(a)}} S^{(a)\dagger}=g_a\Big(\tfrac12,\,\tfrac{i}{2},\,0\Big),
\end{align}
so that
\begin{align}
\big(\nabla S^{(a)\dagger}\times\nabla S^{(a)}\big)=g_a^2\,(0,0,-\tfrac{i}{2}),\nonumber \\
\big(\nabla S^{(a)}\times\nabla S^{(a)\dagger}\big)=g_a^2\,(0,0,+\tfrac{i}{2}).
\end{align}
Dotting with \(\nabla_{\sigma^{(a)}}\tilde\rho=\tfrac14\vec r^{\,(a)}\) gives
\begin{align}
\big(\nabla S^{(a)\dagger}\!\times\!\nabla S^{(a)}\big)\!\cdot\!\nabla\tilde\rho
= -\,\frac{i}{8}\,g_a^2\,r^{(a)}_z,\nonumber \\
\big(\nabla S^{(a)}\!\times\!\nabla S^{(a)\dagger}\big)\!\cdot\!\nabla\tilde\rho
= +\,\frac{i}{8}\,g_a^2\,r^{(a)}_z.
\label{eq:cross-dots}
\end{align}
Using Eq.~\eqref{eq:assoc-sigma} and Eq.~\eqref{eq:cross-dots} with
\(a=1\) (and \(\chi^{(1)}=\kappa_1\sigma^{(1)}_z\)), we define 
\begin{equation}
C^{(1)}_0:=-\frac{i\hbar^3 g_1^2}{48}.
\end{equation}
Then the six associators in Eq.~\eqref{eq:na-master} evaluate to
\begin{align}
A_1=[S^\dagger,S,\rho] \;&=\; -\,C^{(1)}_0\, r^{(1)}_{z}\;\kappa_1\,\sigma^{(1)}_z\!\otimes I,\\
A_2=[S,S^\dagger,\rho] \;&=\; +\,C^{(1)}_0\, r^{(1)}_{z}\;\kappa_1\,\sigma^{(1)}_z\!\otimes I,\\
A_3=[S,\rho,S^\dagger]\;&=\; -\,C^{(1)}_0\, r^{(1)}_{z}\;\kappa_1\,\sigma^{(1)}_z\!\otimes I,\\
A_4=[S^\dagger,\rho,S]\;&=\; +\,C^{(1)}_0\, r^{(1)}_{z}\;\kappa_1\,\sigma^{(1)}_z\!\otimes I,\\
A_5=[\rho,S,S^\dagger]\;&=\; +\,C^{(1)}_0\, r^{(1)}_{z}\;\kappa_1\,\sigma^{(1)}_z\!\otimes I,\\
A_6=[\rho,S^\dagger,S]\;&=\; -\,C^{(1)}_0\, r^{(1)}_{z}\;\kappa_1\,\sigma^{(1)}_z\!\otimes I.
\end{align}
Equivalently, \(A_{1,3,6}=-X^{(1)}\) and \(A_{2,4,5}=+X^{(1)}\) with
\begin{equation}
X^{(1)}:=C^{(1)}_0\, r^{(1)}_{z}\;\kappa_1\,\sigma^{(1)}_z\!\otimes I.
\end{equation}

Similarly, defining \(C^{(2)}_0=-i\hbar^3 g_2^2/48\),
\begin{align}
A_1 \;&=\; -\,C^{(2)}_0\, r^{(2)}_{z}\; I\!\otimes\!\kappa_2\,\sigma^{(2)}_z, & \\
A_2 \;&=\; +\,C^{(2)}_0\, r^{(2)}_{z}\; I\!\otimes\!\kappa_2\,\sigma^{(2)}_z,\\
A_3 \;&=\; -\,C^{(2)}_0\, r^{(2)}_{z}\; I\!\otimes\!\kappa_2\,\sigma^{(2)}_z, & \\
A_4 \;&=\; +\,C^{(2)}_0\, r^{(2)}_{z}\; I\!\otimes\!\kappa_2\,\sigma^{(2)}_z,\\
A_5 \;&=\; +\,C^{(2)}_0\, r^{(2)}_{z}\; I\!\otimes\!\kappa_2\,\sigma^{(2)}_z, & \\
A_6 \;&=\; -\,C^{(2)}_0\, r^{(2)}_{z}\; I\!\otimes\!\kappa_2\,\sigma^{(2)}_z,
\end{align}
or \(A_{1,3,6}=-X^{(2)}\), \(A_{2,4,5}=+X^{(2)}\) with
\(X^{(2)}:=C^{(2)}_0\, r^{(2)}_z\,I\!\otimes\!\kappa_2\sigma^{(2)}_z\).

We now apply the general master equation~\eqref{eq:na-master} with
\begin{equation}
H_S \equiv H_{\mathrm{TFIM}},\qquad
S_1 = g_1 \,\sigma_-^{(1)} \otimes \mathbb{I}, \qquad
S_2 = g_2 \,\mathbb{I} \otimes \sigma_-^{(2)} .
\end{equation}
The complete generator for the two-qubit system coupled to independent
baths has the structure

\begin{widetext}
\begin{align}
\frac{d\rho}{dt}
&= -\frac{i}{\hbar}\,[H_{\mathrm{TFIM}},\rho]
+\sum_{a=1}^2 \Bigg\{
\Gamma_+^{(a)}\Big(S_a \rho S_a^\dagger - \tfrac12\{S_a^\dagger S_a,\rho\}\Big)
+\Gamma_-^{(a)}\Big(S_a^\dagger \rho S_a - \tfrac12\{S_a S_a^\dagger,\rho\}\Big)
\Bigg\} \notag\\
&\quad + \sum_{a=1}^2 \sum_{j=1}^6 \Lambda_j^{(a)}\, \mathcal{A}_j^{(a)}[\rho] ,
\label{eq:TFIMmaster}
\end{align}
\end{widetext}
where $\Gamma_\pm^{(a)}$ and $\Lambda_j^{(a)}$ are determined by the
corresponding bath spectral densities and correlators.
At zero temperature the absorption channels are frozen, i.e.
$\Gamma_-^{(a)}=0$ and $K_-^{(a)}=0$.
Therefore, only $\Lambda_{1,3,6}^{(a)}$ survive, and
Eq.~\eqref{eq:TFIMmaster} simplifies to

\begin{widetext}
\begin{equation}
\frac{d\rho}{dt}
= -\frac{i}{\hbar}\,[H_{\mathrm{TFIM}},\rho]
+\sum_{a=1}^2 \Gamma_+^{(a)}\Big(S_a \rho S_a^\dagger - \tfrac12\{S_a^\dagger S_a,\rho\}\Big)
+\sum_{a=1}^2 \Big(K_+^{(a)} - (K_+^{(a)})^*\Big)\,
\Big([S_a^\dagger,S_a,\rho]+[S_a,\rho,S_a^\dagger]+[\rho,S_a^\dagger,S_a]\Big),
\label{eq:TFIMT0}
\end{equation}
\end{widetext}
which is the direct analogue of Eq.~\eqref{eq:final-N-T0} for the
two-qubit TFIM.

Using the associator evaluations above, the three-term combination in
Eq.~\eqref{eq:TFIMT0} reduces to $-3X^{(a)}$, with
\begin{align}
X^{(1)} = C_0^{(1)} r_z^{(1)} \kappa_1 \,\sigma_z^{(1)} \otimes \mathbb{I},\nonumber \\
X^{(2)} = C_0^{(2)} r_z^{(2)} \kappa_2 \,\mathbb{I} \otimes \sigma_z^{(2)},
\end{align}
where $C_0^{(a)} = -i\hbar^3 g_a^2/48$ and
$r_z^{(a)}=\mathrm{Tr}(\rho\sigma_z^{(a)})$.
Collecting everything, the final nonassociative correction at $T=0$
takes the compact form
\begin{equation}
\mathcal{N}[\rho] = -3i \sum_{a=1}^2 \varepsilon_+^{(a)}\,X^{(a)} ,
\end{equation}
with $\varepsilon_+^{(a)}$ the dispersive (Lamb-shift-like)
contributions associated with each bath.
Substituting $C^{(a)}_0 = -i\hbar^3 g_a^2/48$, the feedback term can
be written compactly as
\begin{equation}
\mathcal{N}[\rho] =
\sum_{a=1}^{2}
\left(-\frac{\hbar^3 g_a^2}{16}\,\varepsilon_{+}^{(a)}\,\kappa_a\right)
r_z^{(a)}\,\sigma_z^{(a)},
\label{eq:NA-final}
\end{equation}
where $r_z^{(a)} = \mathrm{Tr}(\rho\,\sigma_z^{(a)})$.

Two structural remarks are in order.
First, $\mathcal{N}[\rho]$ as given above is an operator
($\sigma_z^{(a)}$ with a state-dependent scalar prefactor) rather than
a superoperator: $\rho$ appears only through the scalar $r_z^{(a)}$,
and the right-hand side does not contain $\rho$ explicitly.
The state-dependence is entirely encoded in the scalar $r_z^{(a)}$.
Second, trace preservation follows from
$\mathrm{Tr}\{\sigma_z^{(a)}\} = 0$, and Hermiticity preservation from
the reality of the coefficient and the Hermiticity of $\sigma_z^{(a)}$.

The complete nonlinear master equation is therefore
\begin{align}
\dot\rho
= -\frac{i}{\hbar}\,[H_{\mathrm{TFIM}},\rho]
+ \sum_{a=1}^2 \Gamma_+^{(a)}\mathcal{D}[\sigma_-^{(a)}]\rho
+ \mathcal{N}[\rho],
\label{eq:ME-nonlinear}
\end{align}
where $\mathcal{N}[\rho]$ is given by Eq.~\eqref{eq:NA-final}.
This nonlinear term couples the instantaneous expectation value
$r_z^{(a)}$ back into the evolution and acts as a self-consistent
longitudinal field along the $z$ direction.


\section{Results}
\label{sec:results}


\subsection{Numerical setup}
\label{sec:numerical_setup}

The Born--Markov master equation is derived under a weak
system--bath coupling assumption, which requires $g \ll \omega_S$ and
$\Gamma_+,\,\varepsilon_+ \ll \omega_S$, where $\omega_S \sim J,\,h$
sets the intrinsic system energy scale.
Since all bath-induced coefficients arise at the same perturbative
order---$\Gamma_+ \propto g^2$, $\varepsilon_+ \propto g^2$,
$\lambda \propto g^2\kappa$---choosing $g \sim J$ or $\Gamma_+ \sim J$
would place the simulation outside the controlled validity regime of
the derivation.
We therefore adopt the parameter set
\begin{equation}
    g_1 = g_2 = 0.2, \quad
    \Gamma^{(a)}_+ = 0.05, \quad
    \varepsilon^{(a)}_+ = 0.01, \quad
    J = 1,
    \label{eq:weak_coupling_params}
\end{equation}
which satisfies $g/J = 0.2$, $\Gamma_+/J = 0.05$, and
$\varepsilon_+/J = 0.01$, consistent with the weak-coupling hierarchy.
The nonassociativity parameters are taken as
$\kappa_1 = \kappa_2 = \kappa$ with
$\kappa \in \{0,\,50,\,100,\,150,\,200\}$,
corresponding to the dimensionless ratio
\begin{equation}
    \frac{\lambda}{\Gamma} \;\equiv\;
    \frac{g^2}{16}\,\frac{\varepsilon_+}{\Gamma_+}\,\kappa
    \;\in\; \{0,\;0.025,\;0.05,\;0.075,\;0.10\}.
\end{equation}
This parametrization ensures that $\mathcal{N}[\rho]$ remains a
controlled perturbation relative to both the coherent and dissipative
scales.

\begin{figure}[t]
  \centering
  \includegraphics[width=\columnwidth]{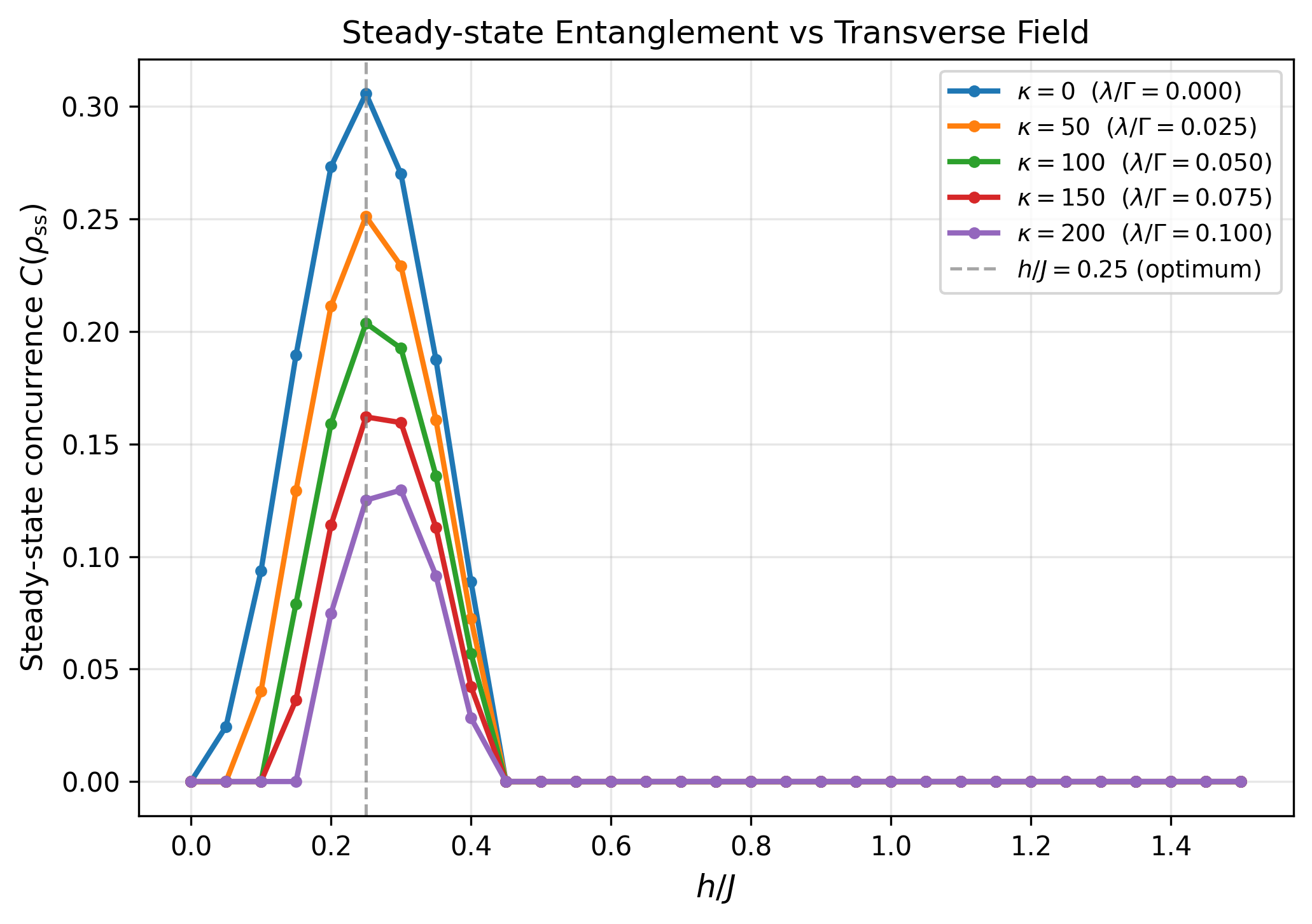}
  \caption{Steady-state concurrence $C(\rho_\mathrm{ss})$ as a
  function of the transverse-field ratio $h/J$ for
  $\kappa \in \{0,\,50,\,100,\,150,\,200\}$
  ($\lambda/\Gamma \in \{0,\,0.025,\,0.05,\,0.075,\,0.10\}$).
  Parameters: $J = 1$, $g = 0.2$, $\Gamma_+ = 0.05$,
  $\varepsilon_+ = 0.01$, $T = 0$;
  initial state
  $|{+}\rangle\langle{+}|\otimes|{+}\rangle\langle{+}|$.
  Steady states are evaluated at $t = 200\,\Gamma_+^{-1}$
  using RK4 with $\Delta t = 0.05$.
  In the weak-coupling regime ($\Gamma_+/J = 0.05$) the
  concurrence is maximized at $h/J \approx 0.25$, where the
  Ising coupling and transverse field are commensurate under the
  given dissipation.
  This optimum lies at smaller $h/J$ than in the strong-coupling
  case ($\Gamma_+/J \sim 1$, where the peak appears near
  $h/J \approx 1.1$), reflecting the reduced weight of
  transverse-field fluctuations when dissipation is weak.
  Nonassociativity monotonically suppresses the peak concurrence
  across the full $h/J$ range; steady-state entanglement
  vanishes for $h/J \gtrsim 0.45$ regardless of $\kappa$.
  The dashed vertical line marks $h/J = 0.25$, used in all
  dynamical simulations.}
  \label{fig:ss_concurrence}
\end{figure}

In this weak-coupling regime, the transverse field is set to
$h = 0.25\,J$, which maximizes the steady-state concurrence for
$\kappa = 0$ (see Fig.~\ref{fig:ss_concurrence}).
This optimum reflects the competition between the Ising coupling $J$,
which tends to lock qubits into product states
$|\!\uparrow\uparrow\rangle$ or $|\!\downarrow\downarrow\rangle$, and
the transverse field $h$, which generates coherent spin flips and
sustains quantum correlations.
At $h/J \approx 0.25$ these two effects are balanced in a way that
maximizes steady-state entanglement under the given dissipation; for
larger $h/J$ the transverse field dominates and destroys the
Ising-mediated correlations.
Setting $h$ at this optimal value ensures that the nonassociative
effect on concurrence is as large as possible and most clearly
observable.
The full $h/J$ dependence is examined separately in
Fig.~\ref{fig:ss_concurrence}.

The system is initialized in the product state
$\rho(0) = |{+}\rangle\langle{+}|\otimes|{+}\rangle\langle{+}|$,
where $|{+}\rangle = (|0\rangle+|1\rangle)/\sqrt{2}$.
Time integration uses a fourth-order Runge--Kutta scheme with step
size $\Delta t = 0.05$; time is measured in units of $\Gamma_+^{-1}$.
Standard Lindblad solvers are not directly applicable because
$\mathcal{N}[\rho]$ depends on the instantaneous expectation values
$r^{(a)}_z = \mathrm{Tr}(\rho\,\sigma^{(a)}_z)$, which must be
recomputed at each Runge--Kutta substep to correctly capture the
nonlinear feedback.
Hermiticity and trace normalization of $\rho$ are enforced after each
time step.
Complete positivity is verified numerically: the minimum eigenvalue of
$\rho(t)$ remains at the level of floating-point precision
(${\sim}10^{-18}$) for all $\kappa$ values explored, confirming that
no physical CP violation occurs.
Because the nonlinear structure of Eq.~\eqref{eq:ME-nonlinear} prevents
a direct application of the linear CP proof given in
Appendix~\ref{app:CP}, this numerical verification serves as the
primary evidence that the nonlinear generator preserves complete
positivity throughout the parameter range explored.


\subsection{Relaxation and coherence dynamics}

\begin{figure*}[htbp]
\centering
\includegraphics[width=0.85\textwidth]{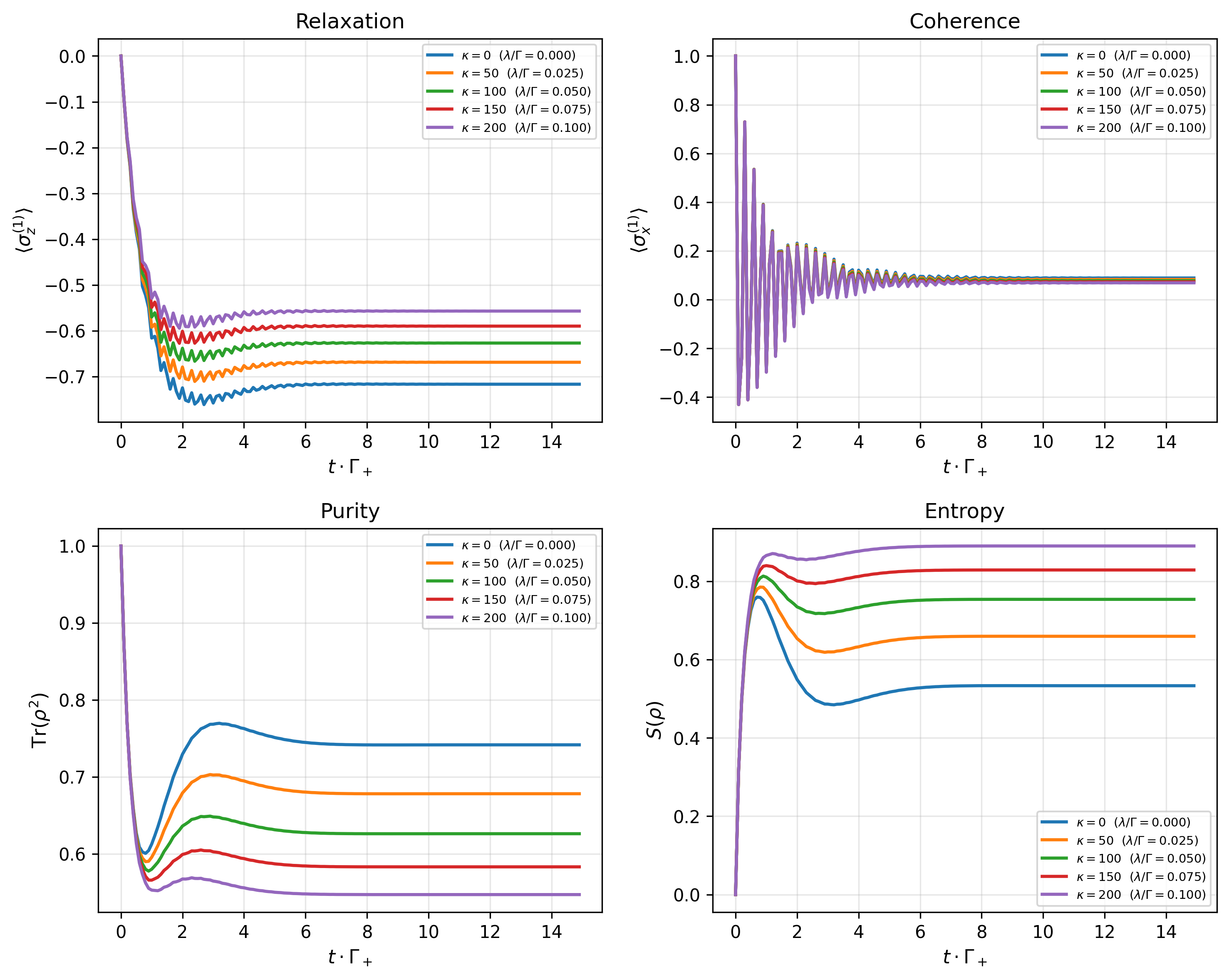}

\caption{Decoherence dynamics of the two-qubit
transverse-field Ising model under increasing nonassociativity
parameter $\kappa \in \{0,\,50,\,100,\,150,\,200\}$, corresponding to
$\lambda/\Gamma \in \{0,\,0.025,\,0.05,\,0.075,\,0.10\}$.
Each panel shows a key observable as a function of dimensionless time
$t\cdot\Gamma_+$:
(a)~population imbalance $\langle\sigma_z^{(1)}\rangle$,
(b)~phase coherence $\langle\sigma_x^{(1)}\rangle$,
(c)~purity $\mathrm{Tr}(\rho^2)$, and
(d)~von Neumann entropy $S(\rho)$.
\textbf{Numerical parameters:}
Ising coupling $J = 1$;
transverse fields $h_1 = h_2 = 0.25\,J$;
coupling constants $g_1 = g_2 = 0.2$;
dispersive contributions
$\varepsilon_+^{(1)} = \varepsilon_+^{(2)} = 0.01$;
amplitude-damping rates
$\Gamma_+^{(1)} = \Gamma_+^{(2)} = 0.05$;
bath temperature $T = 0$ (vacuum limit, $K_-^{(a)} = 0$).
These parameters satisfy the weak-coupling hierarchy
$g/J = 0.2$, $\Gamma_+/J = 0.05$, $\varepsilon_+/J = 0.01$.
\textbf{Initial state:}
$\rho(0)=|{+}\rangle\langle{+}|\otimes|{+}\rangle\langle{+}|$,
$|{+}\rangle = (|0\rangle+|1\rangle)/\sqrt{2}$.
Time integration: fourth-order Runge--Kutta, $\Delta t = 0.05$;
time in units of $\Gamma_+^{-1}$.
As $\kappa$ increases, coherence amplitudes are reduced, purity
decreases, and entropy rises to larger asymptotic values, reflecting
the population-dependent coherent feedback generated by
$\mathcal{N}[\rho]\propto r^{(a)}_z\sigma^{(a)}_z$.
The relaxation timescale in panel~(a) is independent of $\kappa$,
consistent with the purely dispersive character of the nonassociative
correction.}
\label{fig:dynamics}
\end{figure*}

Figure~\ref{fig:dynamics} presents the time evolution of several
observables for increasing values of $\kappa$: the population
imbalance $\langle\sigma_z^{(1)}\rangle$, the phase coherence
$\langle\sigma_x^{(1)}\rangle$, the purity $\mathrm{Tr}(\rho^2)$, and
the von Neumann entropy $S(\rho)$.
In the associative limit $\kappa=0$, the results reproduce the
conventional Lindblad dynamics, where populations relax toward a
steady state, coherences decay with a rate determined by the
dissipator, and the entropy saturates at a finite value corresponding
to a mixed asymptotic state.

When the nonassociative feedback is introduced, the overall shape of
the relaxation curves remains broadly similar, and the characteristic
timescale of relaxation is not dramatically altered in the parameter
regime we explore.
However, the steady-state values and the transient amplitudes change
systematically with $\kappa$.
In particular, the magnitude of coherence oscillations is reduced as
$\kappa$ increases, and the long-time coherence
$\langle\sigma_x^{(1)}\rangle$ approaches smaller absolute values.
Simultaneously, the steady-state purity decreases and the entropy
rises to higher asymptotic values as $\kappa$ grows.
These effects indicate that nonassociativity introduces a
nonlinear, state-dependent coherent feedback that effectively reduces
the amplitude of coherent oscillations and drives the system toward
more mixed steady states. Since the correction is dispersive and
enters through the Liouvillian part of the generator, it does not
introduce additional dissipative channels but instead modifies the
coherent dynamics in a population-dependent manner.
The characteristic relaxation timescale set by $\Gamma_+$
remains unchanged across all $\kappa$ values explored, as confirmed
by panel~(a): the time to reach steady state is determined by
$\Gamma_+^{-1}$, independent of $\kappa$.

\begin{figure}[t]
  \centering
  \includegraphics[width=\columnwidth]{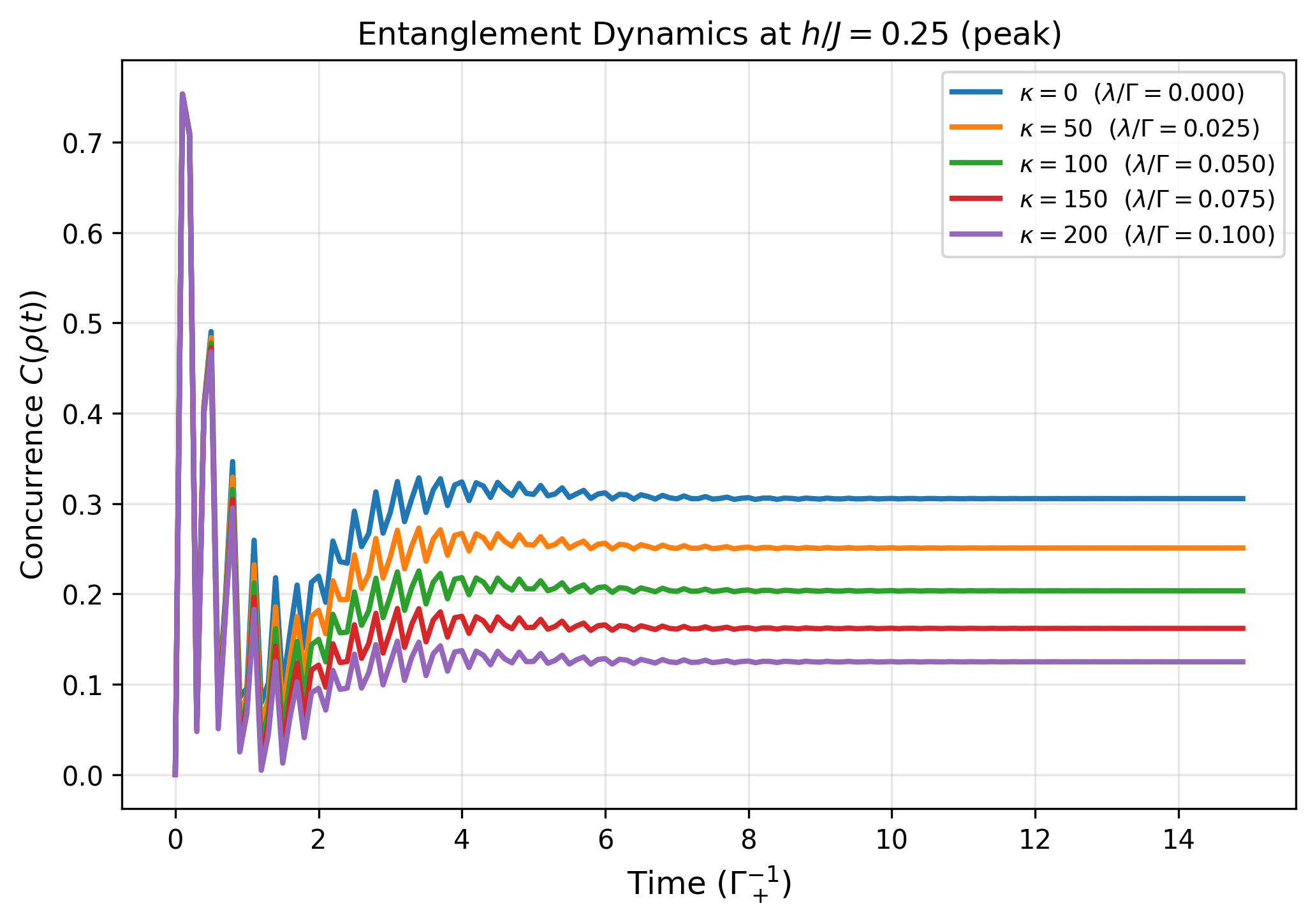}
  \caption{Time evolution of the concurrence $C(\rho(t))$ at
  the optimal transverse field $h/J = 0.25$, where steady-state
  entanglement is maximized in the associative limit $\kappa = 0$
  (cf.\ Fig.~\ref{fig:ss_concurrence}).
  Parameters and initial state are identical to those of
  Fig.~\ref{fig:dynamics}.
  All curves share the same qualitative dynamics: an initial rise
  driven by Ising-mediated correlations, followed by relaxation to
  a nonzero steady state.
  Increasing $\kappa$ systematically reduces the steady-state
  concurrence---from $C_\mathrm{ss} = 0.306$ at $\kappa = 0$ to
  $C_\mathrm{ss} = 0.125$ at $\kappa = 200$---without shifting the
  transient timescale, consistent with the dispersive character of
  $\mathcal{N}[\rho]$.
  The transient peak $C_\mathrm{max}$ varies by less than $0.1\%$
  across all $\kappa$ values, confirming that $\mathcal{N}[\rho]$
  reshapes steady states without accelerating dissipation.}
  \label{fig:entanglement_time}
\end{figure}


\subsection{Correlations and entanglement suppression}

\begin{figure*}[htbp]
\centering
\includegraphics[width=0.85\textwidth]{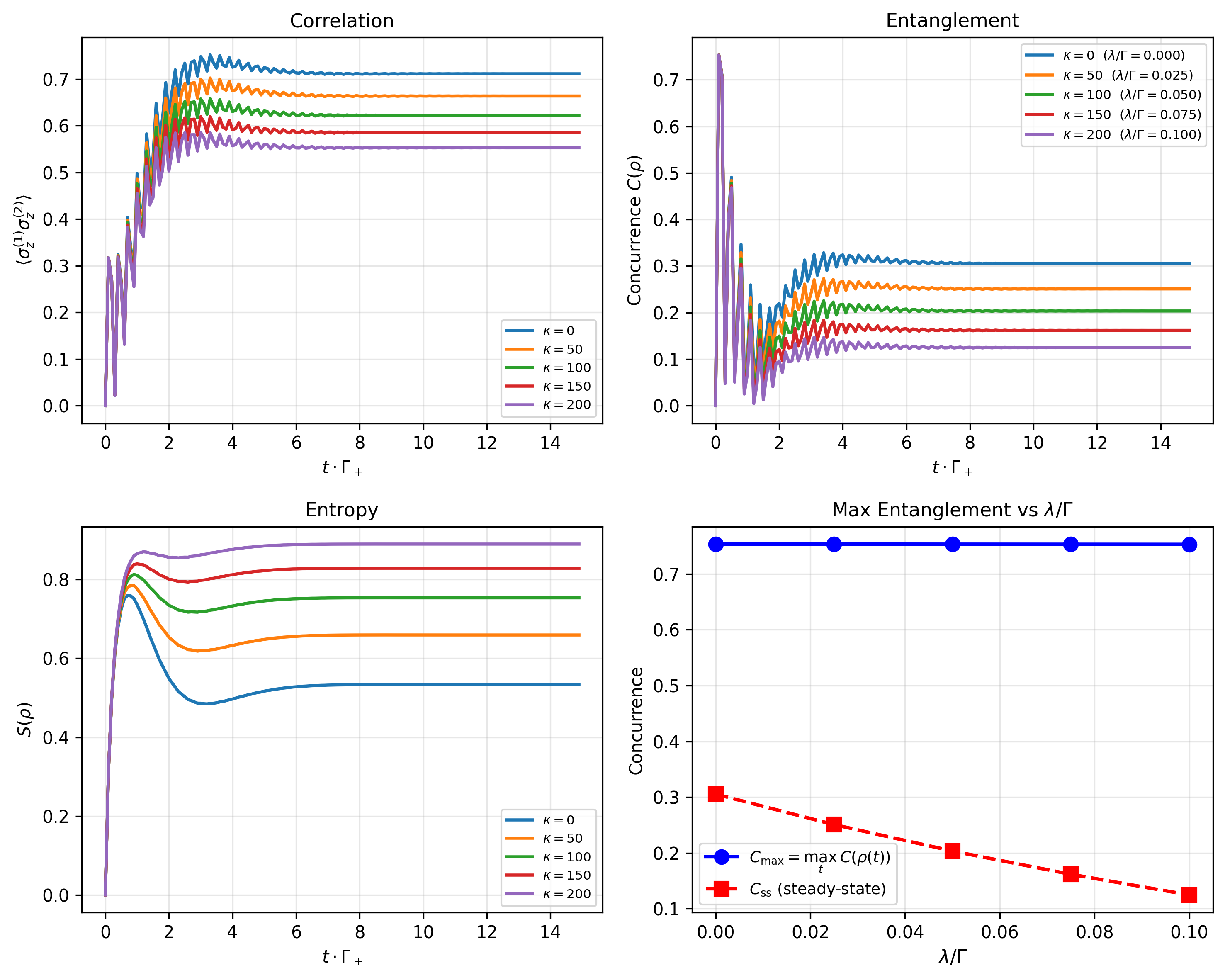}

\caption{Suppression of correlations and entanglement by
nonassociativity for
$\kappa \in \{0,\,50,\,100,\,150,\,200\}$
($\lambda/\Gamma \in \{0,\,0.025,\,0.05,\,0.075,\,0.10\}$).
(a)~Longitudinal spin--spin correlation
$\langle\sigma_z^{(1)}\sigma_z^{(2)}\rangle$,
(b)~concurrence $C(\rho)$,
(c)~von Neumann entropy $S(\rho)$, each as a function of time;
and (d)~maximum concurrence
$C_\mathrm{max}=\max_{t\geq 0}C(\rho(t))$
and steady-state concurrence $C_\mathrm{ss}$ as functions of
$\lambda/\Gamma$.
\textbf{Numerical parameters and initial state} identical to
Fig.~\ref{fig:dynamics}; $h = 0.25\,J$.
As $\kappa$ increases, the steady-state concurrence decreases
monotonically from $0.306$ to $0.125$ (approximately $59\%$
suppression) while the von Neumann entropy increases from $0.533$
to $0.890$, confirming that the nonassociative correction converts
coherent quantum correlations into classical mixtures.
Panel~(d) shows that $C_\mathrm{max}$ (blue circles) remains
nearly constant while $C_\mathrm{ss}$ (red squares) decreases
strongly with $\lambda/\Gamma$, demonstrating that the suppression
is confined to the steady state and originates from dispersive,
state-dependent feedback rather than enhanced dissipation.}
\label{fig:entanglement}
\end{figure*}

The effect of nonassociativity on bipartite correlations and
entanglement is illustrated in Fig.~\ref{fig:entanglement}.
We monitor the longitudinal correlation function
$\langle\sigma_z^{(1)}\sigma_z^{(2)}\rangle$, the concurrence
$C(\rho)$ \cite{Wootters1998}, the entropy $S(\rho)$, and the maximum
concurrence $C_{\mathrm{max}}$ as a function of $\kappa$.
For $\kappa=0$, the exchange coupling $J\sigma_z^{(1)}\sigma_z^{(2)}$
produces a moderate amount of entanglement, which oscillates and
stabilizes at a finite value.
As $\kappa$ increases, the concurrence amplitude systematically
decreases, and the final steady-state correlation becomes weaker.
The monotonic decline of $C_{\mathrm{max}}$ clearly demonstrates that
the nonassociative correction suppresses both the generation and
persistence of quantum entanglement.

Physically, this suppression originates from the nonlinear
coherent feedback generated by the nonassociative correction. The term
proportional to $r_z^{(a)}\sigma_z^{(a)}$ acts as a state-dependent
longitudinal field, producing population-dependent phase shifts between
computational basis states. These nonlinear phase shifts distort the
coherent evolution responsible for entanglement generation and reduce
the build-up of nonlocal correlations. In the presence of the
amplitude-damping dissipator, this modified coherent dynamics leads to
smaller steady-state concurrence and increased entropy, without
introducing an additional dissipative channel.

Comparing steady-state values across increasing $\kappa$, the
concurrence $C(\rho_\mathrm{ss})$ decreases monotonically while the
von Neumann entropy $S(\rho_\mathrm{ss})$ increases, confirming that nonassociativity reduces the build-up of coherent quantum
correlations and shifts the steady state toward more weakly entangled
mixed states. The entropy increase observed here is consistent with the general framework for irreversible entropy production in open quantum systems \cite{Landi2021}, where state-dependent coherent modifications generically lead to increased mixing.
This behavior is consistent with the progressive loss of bipartite
entanglement reported for two-qubit open systems~\cite{YuEberly2004},
\cite{YuEberly2006}, \cite{Cakmak2019} though in the present case, the suppression is
gradual rather than sudden, and originates from an algebraic
deformation of the dynamical generator rather than environmental noise
alone.

A complementary perspective on the entanglement landscape is provided
by scanning the steady-state concurrence as a function of $h/J$ at
fixed dissipation (Fig.~\ref{fig:ss_concurrence}).
The results reveal a pronounced maximum near $h/J \approx 0.25$,
where the interplay between Ising ordering and dissipation is most
favorable for entanglement generation in the weak-coupling regime.
This optimum shifts to smaller $h/J$ compared to the strong-coupling
case, reflecting the reduced role of transverse-field fluctuations
when $\Gamma_+ \ll J$.
Nonassociativity monotonically suppresses this maximum across the full
$h/J$ range, with larger $\kappa$ (larger $\lambda/\Gamma$) producing
stronger suppression.
The time-resolved dynamics at $h/J = 0.25$
(Fig.~\ref{fig:entanglement_time}) confirm that entanglement persists
at long times for all $\kappa \in \{0,\,50,\,100,\,150,\,200\}$
explored here.
Increasing $\kappa$ systematically reduces the steady-state
concurrence from $C_\mathrm{ss} = 0.306$ at $\kappa = 0$ to
$C_\mathrm{ss} = 0.125$ at $\kappa = 200$ ($\lambda/\Gamma = 0.10$),
a suppression of approximately $59\%$, while the transient peak
concurrence $C_\mathrm{max}$ changes by less than $0.1\%$, consistent
with the purely dispersive character of $\mathcal{N}[\rho]$.

\subsection{Discussion: interpretation and physical implications}

The results reported above are obtained in the weak-coupling
regime $g/J = 0.2$, $\Gamma_+/J = 0.05$, consistent with the
Born--Markov derivation.
The nonassociative correction scales as
$\lambda \propto g^2 \varepsilon_+ \kappa$, and within this controlled
regime produces observable modifications to steady states and
entanglement.
The qualitative features---monotonic entanglement suppression, purity
reduction, entropy increase, and timescale invariance---are robust
signatures of the dispersive nonassociative feedback, but the
quantitative magnitudes and the location of the entanglement optimum
in $h/J$ space depend on the ratio $\Gamma_+/J$.
In the weak-coupling limit explored here, the nonassociative correction
produces a $59\%$ reduction in steady-state concurrence at
$\lambda/\Gamma = 0.10$, demonstrating that physically meaningful
signatures persist within the controlled validity regime of the master
equation.

The collective behavior of all observables shows a consistent
picture: increasing $\kappa$ leaves the primary Lindblad relaxation
timescale largely unchanged while modifying the relaxation landscape,
reducing coherence amplitudes, lowering asymptotic purity, increasing
steady-state entropy, and suppressing bipartite entanglement. These
effects originate from the nonlinear coherent feedback generated by
the nonassociative correction, which acts as a state-dependent
longitudinal field proportional to $r_z^{(a)}\sigma_z^{(a)}$.

Importantly, this term does not introduce an additional dissipative
channel. Instead, it modifies the coherent part of the dynamics
through population-dependent energy shifts. In combination with the
existing amplitude-damping processes, this nonlinear feedback alters
the relaxation pathways and leads to modified steady states. In
particular, increasing $\kappa$ leaves residual population in excited
levels and suppresses complete relaxation to the ground state. The
resulting stationary states are therefore generally non-Gibbsian and
represent nonequilibrium steady states stabilized by the nonlinear
feedback.

From a dynamical perspective, the term $N[\rho]\propto r_z\sigma_z$
can be interpreted as a self-consistent longitudinal bias that depends
on the instantaneous state of the system. This feedback-like structure
suggests a natural connection to adaptive control and nonlinear
open-system dynamics. Rather than acting as an additional decoherence
mechanism, the nonassociative correction reshapes the effective energy
landscape and modifies the competition between coherent interactions
and dissipative processes.

This interpretation suggests that the nonassociative generator may
be viewed as an effective nonlinear control resource. Because the final
master equation has the structure of a conventional dissipative model
supplemented by a state-dependent longitudinal field, it can in
principle be emulated in standard associative quantum platforms using
measurement-based feedback or digital-analog simulation. In such an
emulator-level implementation, the instantaneous expectation values
$\langle\sigma_z^{(a)}\rangle$ would be estimated and used to apply a
longitudinal control field proportional to these values, thereby
reproducing the nonlinear generator derived here.

Beyond emulation, the feedback structure also suggests potential
applications in adaptive quantum dynamics. In particular, the
state-dependent longitudinal term naturally produces an effective bias
field
\begin{equation}
h_a^{\mathrm{eff}}(t)=\lambda\, r_z^{(a)}(t),
\end{equation}
which modifies the instantaneous spectrum of the transverse-field
Ising model. Such adaptive biasing is closely related to schemes used
in quantum annealing, where longitudinal fields guide the system toward
low-energy configurations \cite{Mbeng2019}. In the present setting,
the bias emerges intrinsically from the nonlinear dynamics rather than
being externally programmed. This mechanism may influence diabatic
transitions, modify effective gaps, and alter convergence toward target
states.

Finally, the persistence of excited-state population at large
$\kappa$ indicates that the nonassociative feedback can stabilize
metastable nonequilibrium states. Such behavior may be relevant for
applications in dissipative state engineering, nonlinear quantum
annealing, and energy-storage scenarios where incomplete
thermalization and population trapping are advantageous. These
features highlight that nonassociativity acts primarily as a mechanism
for reshaping relaxation pathways rather than accelerating relaxation
itself.


\section{Conclusions}
\label{sec:conclusion}

We have constructed and analyzed a class of nonassociative
open-system dynamics motivated by monopole-twisted phase-space
structures and applied it to a two-qubit transverse-field Ising model.
Starting from a twisted Poisson bivector in momentum space, we
recalled how the Jacobiator generates a nonassociative star product
whose associator is proportional to the divergence of the magnetic
field. Using the Stratonovich--Weyl correspondence on the Bloch sphere,
we then postulated an analogue of this structure for qubit Bloch
variables, adopting an Ising-aligned twist that encodes monopole-like
signatures along the $z$ axis.

On the dynamical side, we derived a general Born--Markov master
equation for a system coupled to a bath when the underlying operator
product is weakly nonassociative. The deviation from associativity
appears at second order via a finite set of operator associators
weighted by bath correlation functions, while the dissipative structure
retains the standard GKLS form. The resulting correction is purely
dispersive and modifies the coherent Liouvillian dynamics without
introducing additional dissipative channels. In the zero-temperature
limit, the correction reduces to a coherent contribution generated by
a combination of three associators of the coupling operators.

Specializing this structure to a two-qubit TFIM with local
amplitude damping, we obtained a nonlinear master equation in which
the instantaneous Bloch-$z$ components feed back into the system as
state-dependent longitudinal fields.
Numerical simulations in the weak-coupling regime
($g/J = 0.2$, $\Gamma_+/J = 0.05$), where the Born--Markov derivation
is quantitatively controlled, show that increasing the
nonassociativity parameter $\kappa$---at fixed dimensionless ratio
$\lambda/\Gamma$ up to $0.10$---reduces coherence amplitudes,
decreases steady-state purity from $0.742$ to $0.547$, increases von
Neumann entropy from $0.533$ to $0.890$, and suppresses steady-state
concurrence by approximately $59\%$, while leaving the characteristic
Lindblad relaxation timescale $\Gamma_+^{-1}$ unchanged.
The transient peak concurrence varies by less than $0.1\%$, confirming
that $\mathcal{N}[\rho]$ reshapes steady states without accelerating
dissipation.
These features arise from population-dependent coherent feedback that
reshapes the relaxation landscape rather than from an additional
decoherence channel.

Conceptually, our analysis demonstrates that algebraic properties
of the underlying operator product---in particular, weak violations of
associativity---can manifest in observable open-system dynamics through
nonlinear, state-dependent coherent corrections. The resulting
generator provides a concrete example of how nonassociativity may be
translated into experimentally accessible signatures at the level of
reduced dynamics. At the same time, because the final master equation
can in principle be emulated by associative measurement-feedback or
digital-analog control schemes, the predicted behavior should be
interpreted as a phenomenological witness of nonassociative dynamics
rather than a unique test of fundamental nonassociativity.

Finally, it is interesting to note that the nonlinear,
state-dependent coherent feedback generated by weak nonassociativity
may be relevant for emerging quantum technology applications. The
population-dependent longitudinal field modifies relaxation pathways
and can stabilize nonequilibrium steady states with residual
excited-state population, suggesting possible relevance for quantum
energy-storage and quantum battery scenarios. Similarly, the adaptive
bias generated by the nonassociative correction resembles the
longitudinal control fields employed in quantum annealing protocols,
where such feedback could influence diabatic transitions and reshape
effective energy landscapes. These observations indicate that, in
analogy with the role of noncommutativity in quantum technologies,
weak nonassociativity may provide a conceptual route toward nonlinear
control resources for open quantum systems.


\section*{Acknowledgments}

This work was supported by the Scientific and Technological Research Council of T\"urkiye (T\"UB\.ITAK) under Project Number 124F109. ESY would like to dedicate this paper to her father for his 58th birthday.


\section*{Data and code availability}

All numerical simulations, data analysis, and figure generation
presented in this work were performed using a custom Python code
developed by the authors. The script \texttt{nonassociative\_simulation.py} reproduces all figures
presented in this work using the weak-coupling parameter set
($g = 0.2$, $\Gamma_+ = 0.05$, $\varepsilon_+ = 0.01$,
$h/J = 0.25$, $J = 1$) and generates the corresponding data files.
Complete positivity is verified within the same script by monitoring
the minimum eigenvalue of $\rho(t)$ at each recorded time step.

\onecolumngrid
\appendix

\section{Nonassociative corrections to the Born--Markov master equation}
\label{app:NAderivation}

In this appendix we provide the detailed derivation of the
nonassociative corrections to the Born--Markov master equation
summarized in Sec.~\ref{sec:generalME}.

\subsection{Interaction picture and second-order expansion}

We start from the Liouville--von Neumann equation in the interaction
picture,
\begin{equation}
\frac{d}{dt} \hat{\rho}_I(t) = -\frac{i \alpha}{\hbar} \left[ \hat{H}_I(t), \, \hat{\rho}_I(t) \right],
\end{equation}
where $\hat{H}_I(t)=e^{t\mathcal{L}_0}H_{SB}$ and $\mathcal{L}_0$ is
defined in Sec.~\ref{sec:generalME}.
Integrating once,
\begin{equation}
\hat{\rho}_I(t) = \hat{\rho}_I(0) - \frac{i \alpha}{\hbar} \int_0^t dt_1 \left[ \hat{H}_I(t_1), \, \hat{\rho}_I(t_1) \right],
\end{equation}
and substituting back yields the second-order expansion
\begin{align}
\frac{d}{dt} \hat{\rho}_I(t) =& -\frac{i \alpha}{\hbar} \left[ \hat{H}_I(t), \, \hat{\rho}_I(0) \right] \nonumber \\ &
- \frac{\alpha^2}{\hbar^2} \left[ \hat{H}_I(t), \, \int_0^t dt_1 \left[ \hat{H}_I(t_1), \, \hat{\rho}_I(t_1) \right] \right].
\end{align}
Tracing over the bath degrees of freedom and assuming that the
first-order term vanishes due to a traceless interaction, one finds
\begin{equation}
\frac{d}{dt} \hat{\rho}_S(t) = -\frac{\alpha^2}{\hbar^2} \, \mathrm{Tr}_B \left\{ \left[ \hat{H}_I(t), \, \int_0^t dt_1 \left[ \hat{H}_I(t_1), \, \hat{\rho}_I(t_1) \right] \right] \right\}.
\end{equation}
Under the Born approximation,
\begin{equation}
\hat{\rho}_I(t_1) \approx \hat{\rho}_S(t) \otimes \hat{\rho}_B,
\end{equation}
and after extending the upper limit to infinity (Markov approximation)
one obtains Eq.~\eqref{eq:von-neu-jacob}.

\subsection{Jacobiator and associator decomposition}

At this stage the form of the kernel still resembles the associative
case.
However, in a nonassociative algebra the nested commutator carries
additional structure.
Using the definition of the Jacobiator and the
identity~\eqref{jabobiator-associator}, the double commutator can be
rewritten as

\begin{align}
 \left[ \hat{H}_I(t), [\hat{H}_I(t'), \hat{\rho}_S \hat{\rho}_B] \right]
 &= -\left[ \hat{H}_I(t'), [\hat{\rho}_S \hat{\rho}_B,\hat{H}_I(t)] \right]
   -\left[ \hat{\rho}_S \hat{\rho}_B, [\hat{H}_I(t), \hat{H}_I(t')] \right]
   -\sum_{j=1}^{6} \mathcal{A}_j(t,t'),
\label{eq:doublec}
\end{align}

where the six associator structures are

\begin{align}
\mathcal{A}_1&=[\hat H(t),\hat H(t'),\hat\rho_S\hat\rho_B]
=\big(\hat H(t)\hat H(t')\big)\hat\rho_S\hat\rho_B-\hat H(t)\big(\hat H(t')(\hat\rho_S\hat\rho_B)\big),
\\
\mathcal{A}_2&=-[\hat H(t),\hat\rho_S\hat\rho_B,\hat H(t')]
=-\Big(\big(\hat H(t)(\hat\rho_S\hat\rho_B)\big)\hat H(t')-\hat H(t)\big((\hat\rho_S\hat\rho_B)\hat H(t')\big)\Big),
\\
\mathcal{A}_3&=[\hat H(t'),\hat\rho_S\hat\rho_B,\hat H(t)]
=\big(\hat H(t')(\hat\rho_S\hat\rho_B)\big)\hat H(t)-\hat H(t')\big((\hat\rho_S\hat\rho_B)\hat H(t)\big),
\\
\mathcal{A}_4&=-[\hat H(t'),\hat H(t),\hat\rho_S\hat\rho_B]
=-\Big(\big(\hat H(t')\hat H(t)\big)\hat\rho_S\hat\rho_B-\hat H(t')\big(\hat H(t)(\hat\rho_S\hat\rho_B)\big)\Big),
\\
\mathcal{A}_5&=[\hat\rho_S\hat\rho_B,\hat H(t),\hat H(t')]
=\big((\hat\rho_S\hat\rho_B)\hat H(t)\big)\hat H(t')-(\hat\rho_S\hat\rho_B)\big(\hat H(t)\hat H(t')\big),
\\
\mathcal{A}_6&=-[\hat\rho_S\hat\rho_B,\hat H(t'),\hat H(t)]
=-\Big(\big((\hat\rho_S\hat\rho_B)\hat H(t')\big)\hat H(t)-(\hat\rho_S\hat\rho_B)\big(\hat H(t')\hat H(t)\big)\Big).
\end{align}

In the associative limit all $\mathcal{A}_j$ vanish and one recovers
the standard Born--Markov kernel.

\subsection{Fermionic bath and correlation functions}

We now specialize to a fermionic bath $B$ with interaction Hamiltonian
\begin{equation}
\hat H(t)=\hbar\big(\hat S\,\hat B^\dagger(t)+\hat S^\dagger\,\hat B(t)\big),
\qquad
\hat B(t)=\sum_k g_k^*\,\hat b_k\,e^{-i\omega_k t},
\ \
\hat B^\dagger(t)=\sum_k g_k\,\hat b_k^\dagger\,e^{+i\omega_k t},
\end{equation}
with $\{\hat b_k,\hat b_{k'}^\dagger\}=\delta_{kk'}$.
The reduced state is $\hat\rho_S(t)$ and the bath is a stationary
state $\hat\rho_B$ at temperature $T$.
Expanding the $\mathcal{A}_j$ in terms of $\hat S$, $\hat S^\dagger$,
$\hat B$, and $\hat B^\dagger$ and tracing over the bath eliminates all
terms with an odd number of bath operators.
The surviving contributions can be expressed through the two-time
correlators
\begin{equation}
C_+(\tau):=\big\langle B(t)\,B^\dagger(t')\big\rangle_B,\qquad
C_-(\tau):=\big\langle B^\dagger(t)\,B(t')\big\rangle_B,
\qquad \tau=t-t',
\end{equation}
where the expectation value is taken in $\rho_B$.

The nonassociative correction to the Born--Markov kernel can then be
written as a sum of six terms,
\begin{equation}
\Delta_j[\rho_S](t) := -\frac{1}{\hbar^2}\int_{0}^{\infty} dt'\;\mathrm{Tr}_B\{-\mathcal{A}_j(t,t')\},
\end{equation}
which evaluate to

\begin{align}
\Delta_{1}[\rho_S](t)&=\int_0^\infty dt' \Big([\hat S^\dagger, \hat S, \rho_S] \langle \hat B(t)\hat B^\dagger(t')\rangle_B + [\hat S, \hat S^\dagger, \rho_S]\langle \hat B^\dagger(t)\hat B(t')\rangle_B\Big),\\
\Delta_{2}[\rho_S](t)&=-\int_0^\infty dt' \Big([\hat S, \rho_S, \hat S^\dagger] \langle \hat B(t')\hat B^\dagger(t)\rangle_B + [\hat S^\dagger, \rho_S, \hat S]\langle \hat B^\dagger(t')\hat B(t)\rangle_B\Big),\\
\Delta_{3}[\rho_S](t)&=\int_0^\infty dt' \Big([\hat S, \rho_S, \hat S^\dagger] \langle \hat B(t)\hat B^\dagger(t')\rangle_B + [\hat S^\dagger, \rho_S, \hat S]\langle \hat B^\dagger(t)\hat B(t')\rangle_B\Big),\\
\Delta_{4}[\rho_S](t)&=-\int_0^\infty dt' \Big([\hat S^\dagger, \hat S, \rho_S] \langle \hat B(t')\hat B^\dagger(t)\rangle_B + [\hat S, \hat S^\dagger, \rho_S]\langle \hat B^\dagger(t')\hat B(t)\rangle_B\Big),\\
\Delta_{5}[\rho_S](t)&=\int_0^\infty dt' \Big([\rho_S, \hat S, \hat S^\dagger]\langle \hat B^\dagger(t)\hat B(t')\rangle_B+ [\rho_S, \hat S^\dagger, \hat S] \langle \hat B(t)\hat B^\dagger(t')\rangle_B\Big),\\
\Delta_{6}[\rho_S](t)&=-\int_0^\infty dt' \Big([\rho_S, \hat S, \hat S^\dagger]\langle \hat B^\dagger(t')\hat B(t)\rangle_B+ [\rho_S, \hat S^\dagger, \hat S] \langle \hat B(t')\hat B^\dagger(t)\rangle_B\Big).
\end{align}

Introducing $\tau=t-t'$ and using stationarity of the bath, the six
contributions can be compactly written as
\begin{align}
\mathcal{N}[\rho_S]
&= \int_0^\infty d\tau\,
\Big\{
\alpha_1(\tau)\,[S^\dagger,S,\rho_S]
+\alpha_2(\tau)\,[S,S^\dagger,\rho_S]
+\alpha_3(\tau)\,[S,\rho_S,S^\dagger]
\nonumber\\[-2pt]
&\hspace{5.1em}
+\alpha_4(\tau)\,[S^\dagger,\rho_S,S]
+\alpha_5(\tau)\,[\rho_S,S,S^\dagger]
+\alpha_6(\tau)\,[\rho_S,S^\dagger,S]
\Big\},
\end{align}
with coefficients
\begin{align}
\alpha_1(\tau) &= C_+(\tau) - C_+(-\tau),
&\qquad
\alpha_2(\tau) &= C_-(\tau) - C_-(-\tau),
\\
\alpha_3(\tau) &= C_+(\tau) - C_+(-\tau),
&\qquad
\alpha_4(\tau) &= C_-(\tau) - C_-(-\tau),
\\
\alpha_5(\tau) &= C_-(\tau) - C_-(-\tau),
&\qquad
\alpha_6(\tau) &= C_+(\tau) - C_+(-\tau).
\end{align}
The bath correlators
\begin{align}
C_+(\tau) &= \sum_k |g_k|^2\, e^{-i\omega_k \tau}\,\big(1-n_F(\omega_k)\big),\\
C_-(\tau) &= \sum_k |g_k|^2\, e^{-i\omega_k \tau}\, n_F(\omega_k),
\end{align}
can be expressed in terms of the spectral density $J(\omega)$ and the
Fermi function $n_F(\omega)$ in the continuum limit.

\subsection{Markov limit and final generator}

In the Markov approximation one introduces the time-integrated kernels
\begin{equation}
K_\pm \;:=\; \int_0^\infty d\tau\, C_\pm(\tau)
\;=\; \frac{\Gamma_\pm + i\,\varepsilon_\pm}{2},
\end{equation}
with dissipative rates and dispersive (Lamb-shift-like) contributions

\begin{align}
\Gamma_\pm &= 2\pi\, J(\omega_\star)\, f_\pm(\omega_\star),
\qquad
\varepsilon_\pm = -2\,\mathcal{P}\!\int_0^\infty d\omega\, \frac{J(\omega)\, f_\pm(\omega)}{\omega-\omega_\star},
\label{eq:GammaEps}
\\
f_+(\omega)&:=1-n_F(\omega),\qquad f_-(\omega):=n_F(\omega),
\end{align}

where $\omega_\star$ denotes the system transition frequency selected
by the coupling.
The nonassociative part of the generator thus acquires the form quoted
in Eq.~\eqref{eq:na-master}, with coefficients $\Lambda_j$ given in
Eq.~\eqref{eq:LambdaChains}.
In the zero-temperature, vacuum-like case, $K_-=0$ and the correction
reduces to Eq.~\eqref{eq:final-N-T0}.

Finally, combining the standard GKLS dissipator with $\Gamma_\pm$ and
the nonassociative correction $\mathcal{N}[\rho_S]$ yields the
complete master equation~\eqref{eq:na-master}, which forms the
starting point for the two-qubit TFIM analysis in Sec.~\ref{sec:model}.

\section{Complete positivity of the generalized master equation}
\label{app:CP}

In this appendix we establish that the \emph{general} master
equation, Eq.~\eqref{eq:na-master}, generates a completely positive
(CP) dynamical map when the secular approximation is applied.
This proof applies to Eq.~\eqref{eq:na-master} in its linear form,
where $\mathcal{N}[\rho_S]$ acts as a linear superoperator on $\rho_S$
through the six associator structures.
It does \emph{not} directly extend to the nonlinear reduced equation,
Eq.~\eqref{eq:ME-nonlinear}, obtained after the Stratonovich--Weyl
mapping; complete positivity for that equation is verified numerically
in Sec.~\ref{sec:numerical_setup}.

\paragraph{Bohr-frequency decomposition.}
The system operator is expanded as
\begin{equation}
S(t)=\sum_\omega e^{-i\omega t}S(\omega),\qquad
S^\dagger(t)=\sum_\omega e^{+i\omega t}S^\dagger(\omega).
\end{equation}
This decomposition is inserted into the interaction Hamiltonian and
hence into the second-order kernel, including the associator terms
$\mathcal{A}_1,\dots,\mathcal{A}_6$.

\paragraph{Secular approximation.}
After the Bohr-frequency expansion, each contribution contains
oscillating prefactors $e^{-i(\omega-\omega')t}$.
Cross terms with $\omega\neq\omega'$ average to zero on the
coarse-grained time scale.
In the secular (rotating-wave) approximation we therefore discard all
such cross terms, retaining only the resonant $\omega=\omega'$
contributions.
This step is applied uniformly both to the standard dissipative kernel
and to the non-associative associator contributions
$\mathcal{A}_1,\dots,\mathcal{A}_6$.

In practice, this means that in each $\mathcal{A}_j$ the operators
$S(t),S^\dagger(t)$ are expanded into Bohr-frequency components.
The resulting double sums over $\omega,\omega'$ produce oscillating
factors $e^{-i(\omega-\omega')t}$.
The secular approximation then discards all $\omega\neq\omega'$ terms,
so that each $\mathcal{A}_j$ reduces to a sum of resonant
contributions with $\omega=\omega'$.

\paragraph{Example: Secular approximation for $\Delta_1[\rho_S]$.}

Starting from the definition,
\begin{equation}
\Delta_1[\rho_S](t) \;=\; -\frac{1}{\hbar^2}\int_0^\infty d\tau\;
\mathrm{Tr}_B\{A_1(t,t-\tau)\},
\end{equation}
with
\begin{equation}
A_1(t,t') \;=\; [S^\dagger(t),\,S(t'),\,\rho_S]\;B(t)B^\dagger(t').
\end{equation}
We expand the system operators as
\begin{equation}
S(t)=\sum_{\omega'} e^{-i\omega' t} S(\omega'),
\qquad
S^\dagger(t)=\sum_{\omega} e^{+i\omega t} S^\dagger(\omega).
\end{equation}
Inserting into $A_1$ gives
\begin{equation}
A_1(t,t-\tau) = \sum_{\omega,\omega'}
e^{+i\omega t}\,e^{-i\omega'(t-\tau)}\;
[\,S^\dagger(\omega),\,S(\omega'),\,\rho_S]\;B(t)B^\dagger(t-\tau).
\end{equation}
Tracing over the bath yields correlation functions $C_+(\tau)$ and
\begin{equation}
\Delta_1[\rho_S](t)
= \sum_{\omega,\omega'} e^{i(\omega-\omega')t}\;
\left( \int_0^\infty d\tau\, e^{+i\omega'\tau} C_+(\tau)\right)\;
[\,S^\dagger(\omega),\,S(\omega'),\,\rho_S].
\end{equation}
We define
\begin{equation}
\Gamma_+(\omega') := \int_0^\infty d\tau\, e^{+i\omega'\tau}\,C_+(\tau).
\end{equation}
The prefactor $e^{i(\omega-\omega')t}$ oscillates rapidly for
$\omega\neq\omega'$ and averages to zero on the coarse-grained time
scale. In the secular approximation we therefore retain only
$\omega=\omega'$, leading to
\begin{equation}
\Delta_1[\rho_S](t)
= \sum_{\omega} \Gamma_+(\omega)\;
[\,S^\dagger(\omega),\,S(\omega),\,\rho_S].
\end{equation}

Thus the contribution of $A_1$ reduces to a resonant sum with positive
rates $\Gamma_+(\omega)$, consistent with the GKLS structure.
After the secular approximation the dissipative part of the generator
has the standard GKLS structure with positive rates $\Gamma_\pm(\omega)$.
This ensures complete positivity.

\paragraph{Associator contributions.}
The additional non-associative terms combine into $\mathcal{N}[\rho_S]$.
Two structural properties hold:
(i) By the 3-cyclic trace,
$\mathrm{Tr}_S\{\mathcal{N}[\rho_S]\}=0$, so trace preservation is
automatic.
(ii) The integrated coefficients $\Lambda_j$ are purely imaginary, so
$\mathcal{N}$ is Hermitian and maps Hermitian operators to Hermitian
operators.
After the secular approximation only resonant components remain, which
act as dispersive (Lamb-shift-like) corrections to the Hamiltonian
part of the generator.

Consequently, the dynamics generated by the general master
equation Eq.~\eqref{eq:na-master} is trace-preserving,
Hermiticity-preserving, and completely positive \cite{Vacchini2000}.
This conclusion holds for the linear generator; the nonlinear
equation Eq.~\eqref{eq:ME-nonlinear} satisfies these properties within
the numerical precision of the simulations, as confirmed by the
eigenvalue monitoring described in Sec.~\ref{sec:numerical_setup}.

\bibliographystyle{unsrt}
\bibliography{references}

\end{document}